\documentclass[11pt]{article}
\usepackage{hyperref}
\usepackage{amsmath}
\usepackage{subfigure}
\usepackage{epsfig}

\begin{document}

\title{\bf Invisibility cloak without singularity}

\author{Wei Xiang Jiang, Tie Jun Cui\footnote{Electronic mail: tjcui@seu.edu.cn}, Xin Mi Yang, Qiang Cheng\\
{\small\it State Key Laboratory of Millimeter Waves}  \\
{\small\it Institute of Target Characteristics and Identification,
Department of Radio Engineering} \\ {\small\it Southeast University, Nanjing 210096, P. R. China.} \\
Ruopeng Liu, David R. Smith\footnote{Electronic mail:
drsmith@ee.duke.edu}\\
{\small\it Department of Electrical and Computer Engineering} \\
{\small\it Duke University, Box 90291, Durham, North Carolina 27708,
USA.}}

\date{}
\maketitle

\begin{abstract}

An elliptical invisible cloak is proposed using a coordinate
transformation in the elliptical-cylindrical coordinate system,
which crushes the cloaked object to a line segment instead of a
point. The elliptical cloak is reduced to a nearly-circular cloak if
the elliptical focus becomes very small. The advantage of the
proposed invisibility cloak is that none of the parameters is
singular and the changing range of all parameters is relatively
small.

\vskip 5mm


\vskip 5mm


\vskip 5mm \noindent
\end{abstract}

\newpage

Great attention has been paid to the electromagnetic (EM) cloaks due
to the exciting property of invisibility [1-15]. The coordinate
transformation method to control the EM fields was reported in Ref.
[1]. The cloaking principle was experimentally demonstrated using
reduced constitutive parameters in the microwave regime [2]. Many
further theoretical and numerical studies have been devoted in the
past two years [3-11]. A big problem in the full-parameter cloak is
that some parameter components exist singularity on the inner
boundary of the cloak, which makes the full cloak difficult to
achieve even using metamaterials [2]. Recently, the elliptical
cloaks have been designed and investigated in different coordinate
systems [12-14]. However, the cloak parameters are fully anisotropic
and some components still have singular points on the inner
boundary. In principle, the above circular and elliptical cloaks try
to crush the cloaked objects to a point, which results in the
singular parameters.

In view of the difficulty to realize the full-parameter cloak and
the imperfection of reduced-parameter cloak, a recently published
theory has suggested a carpet cloak, which can hide any objects
under a metamaterial carpet [15]. Different from the completely
invisible cloak, the carpet cloak crushes the hidden object to a
conducting sheet. The great advantage of the carpet cloak is that it
does not require singular values for the material parameters.
However, the carpet cloak can only hide objects placed under a
conducting plane, and cannot hide objects in free space.

In this work, we propose an elliptical invisible cloak using the
coordinate transformation in the classical elliptical-cylindrical
coordinate system. Instead of shrinking the cloaked object to a
point, the proposed cloak crushes the object to a line segment,
which avoids any singularities in the constitutive parameters.
Closed-form formulations are derived for both permittivity and
permeability tensors. If the focus of the ellipse is very small, the
elliptical cloak approaches a circular cloak. The advantage of the
proposed invisibility cloak is that none of the parameters is
singular and the changing range of all parameters is relatively
small.

We consider the elliptical-cylindrical cloak in two dimensions. Due
to the elliptical shape of the cloak, we construct the coordinate
transformation in the classical elliptically-cylindrical coordinate
system $(\xi,~\eta,~z)$, whose relationship to the Cartesian
coordinates $(x,~y,~z)$ is written as
\begin{eqnarray}
x=p\cosh \xi \cos \eta,\quad y=p \sinh \xi \sin \eta,\quad z=z,~~~~
\end{eqnarray}
in which $2p$ is the focus of the ellipse, as shown in Fig. 1. In
the elliptical cylindrical coordinate system, if we assume $p$ to be
constant, then isolines for $\xi$ can be a series of elliptical
cylindrical shells with the same focus value. Then a spatial
transformation from the elliptical region $\xi\in[0, \xi_2]$ to the
annular region $\xi'\in[\xi_1, \xi_2]$ can be represented
mathematically as
\begin{eqnarray}
\xi'=\frac{\xi_2-\xi_1}{\xi_2}\xi+\xi_1,~\eta'=\eta,~z'=z,
\end{eqnarray}
where $\xi_1$ and $\xi_2$ are coordinate parameters of the inner and
outer boundaries of the elliptical cloak. Let $a_1$ and $a_2$ denote
the lengths of major axes for inner and outer shells of the cloak.
The nonlinear relationship between coordinate parameters and the
lengths of major axis can be expressed as
$\xi_i=\ln\big(a_i/p+\sqrt{(a_i/p)^2-1}\big)$, $i=1,~2$. We remark
that the inner and outer ellipses are of the same focus value $2p$.
Hence the inner boundary will be crushed to the line segment $2p$
using the coordinate transformation.

Similar to the procedure stated in Ref. [1], one can deduce the
parameter tensors of the elliptical cloak. The relative permittivity
and permeability tensors are expressed as
\begin{eqnarray}
&&\varepsilon'_{\xi'}=\mu'_{\xi'}=\frac{\xi_2-\xi_1}{\xi_2},\\
&&\varepsilon'_{\eta'}=\mu'_{\eta'}=\frac{\xi_2}{\xi_2-\xi_1},\\
&&\varepsilon'_{z'}=\mu'_{z'}=\frac{\xi_2}{\xi_2-\xi_1}\frac{\cosh^2\beta-\cos^2\eta'}{\cosh^2\xi'-\cos^2\eta'},
\end{eqnarray}
in which $\beta=\xi_2(\xi'-\xi_1)/(\xi_2-\xi_1)$, $\xi_1\leq\xi'\leq
\xi_2$, and $0\leq \eta'\leq 2\pi$.

Eqs. (3)-(5) provide full design parameters for the elliptical cloak
in the classical elliptically-cylindrical coordinates. Clearly, the
cloak is composed of inhomogeneous and uniaxially anisotropic
metamaterials. For circularly cylindrical cloaks with full
parameters, singular material parameters are distributed on the
inner boundary [4-11], which are difficult to realize in the actual
applications [2]. For elliptical cloaks which shrink the cloaked
objects to a point, singular values still exist on the inner
boundary of the cloaks [12-14]. The material parameters for the
proposed elliptical-cylindrical cloak which crushes the cloaked
object to the line segment $2p$, however, have no singularity. This
makes it possible to realize the full-parameter cloak using
metamaterials.

In order to validate the elliptical cloak with the designed
parameters, we make full-wave simulations based on the method of
finite elements. Either TE-polarized or TM-polarized time-harmonic
incident plane waves can be used. In the case of TE polarization,
only $\mu_{\xi}$, $\mu_{\eta}$, and $\varepsilon_{z}$ components of
the material parameters are required for the simulations; in the
case of TM polarization, however, we only need $\varepsilon_{\xi}$,
$\varepsilon_{\eta}$, and $\mu_{z}$ components. In this work, we
only consider the TM case for space reason. The working frequency is
chosen as 9 GHz.

The example we considered is a perfectly electrical conducting (PEC)
cylinder covered by the elliptical cloak, in which the lengths of
major axes for inner and outer ellipses are 0.025 m and 0.05 m,
respectively, and half of the focus length is $p=0.015$ m. We show
that the magnetic fields are smoothly excluded from the interior
region with the minimal scattering in any directions when the
incident direction of plane waves is parallel to the long axis of
the ellipse. When the incident waves are perpendicular to the long
axis, however, a big scattering outside the cloaking region is
observed. The physical reason for the incident-angle dependence is
that the cloaked object is crushed to the line segment $2p$, instead
of a point. In the first case, the incident direction is parallel to
the line segment, which produces the minimum scattering; in the
second case, the incident direction is vertical to the line segment,
which produces the maximum scattering.

When the focus length $2p$ of the elliptical cloak approach zero,
however, the line segment $2p$ would be shrink to a point. As a
result, the elliptical cloak would approach a circular cloak, and
the cloaking effect would become much better. Nevertheless, $p$
cannot be zero due to the elliptical coordinate transformation (1).
Hence we select $p$ as a small value.

When $p$ is chosen as a small value, the elliptical cloak is nearly
a circular cloak with outer radius $R_2$ and inner radius $R_1$. In
such a case, $\varepsilon_{\xi}$ becomes $\varepsilon_{r}$ and
$\varepsilon_{\eta}$ becomes $\varepsilon_{\phi}$, indicating the
radian and angular components of the permittivity. In order to
compare with the ordinary circular cloak [4-11], in the following
simulations, we choose the shape of cloak as \textbf{an exact
circle} while the material parameters are given by:
\begin{eqnarray}
&&\varepsilon'_{r}=\mu'_{r}=k,\\
&&\varepsilon'_{\phi}=\mu'_{\phi}=\frac{1}{k},\\
&&\varepsilon'_{z}=\mu'_{z}=\frac{\cosh^2\beta}{k\cosh^2\xi},
\end{eqnarray}
in which $k=(\xi_2-\xi_1)/\xi_2$, $\beta=(\xi-\xi_1)/k$,
$\xi=\ln\big(r/p+\sqrt{(r/p)^2-1}\big)$,
$\xi_i=\ln\big(R_i/p+\sqrt{(R_i/p)^2-1}\big)$, $i=1,~2$.

The full-wave simulation result for a circular TM cloak with
$p=0.001$ m is illustrated in Fig. 2(a). We observe that the phase
fronts are bent smoothly around the PEC object inside the cloak, and
the fields are smoothly excluded from the interior region with the
tiny scattering. Similar phenomena are observed for other small
values of $p$, for example, $p=0.003$ m and 0.005 m (not shown).

All material parameters for above circular cloaks have relatively
small ranges. Figure 2(b) illustrates the parameter distributions
inside the TM cloaks when $p=0.001$ m, 0.003 m, and 0.005 m.
Obviously, for each $p$, $\varepsilon_{r}$ and $\varepsilon_{\phi}$
are constants, and $\mu_{z}$ ranges from 0 to a small constant, all
of which can be realized using metamaterials. From Fig. 2(a), we
clearly observe that the proposed cloak with smaller $p$ can achieve
better cloaking performance, because the cloaked object is crushed
nearly to a point, which results in tiny scattered fields.

In the earlier example, the material parameters are gradiently
distributed in the circular region, which are difficult to implement
in real experiments. In order to design a practical cloak, we
consider a layered homogeneous circular cloak which is divided into
8 layers. When we choose $p=0.001$ m, the permittivity components in
all layers are constants: $\varepsilon_{r}=0.151$ and
$\varepsilon_{\phi}=6.641$, while the permeability component
$\mu_{z}$ is given by 0.025, 0.193, 0.466, 1.063, 1.90, 2.756, 3.95,
and 5.60 in all layers. We remark that we choose different
thicknesses for the 8 layers due to the nonlinear distribution of
the permeability.

Figure 3(a) illustrates the simulation results of the layered cloak.
We clearly observe a good performance of the invisible cloaking
although we applied the designed parameters for a nearly-circular
cloak to a layered circular cloak. From Fig. 3(a), the cloak forces
the incoming plane waves to propagate around the inner cloaked
region, and such waves return to their original propagation
directions without distorting the waves outside the cloak.

In real applications, artificial metamaterials are always lossy.
Hence it is important to study the lossy effect of cloak on the
invisible property. Figure 3(b) shows the magnetic-field
distributions for the cloak with electric and magnetic-loss tangents
of 0.03. From this figure, although the inhomogeneous cloak has been
partitioned into eight homogeneous layers and both the permittivity
and permeability have a relatively large loss, we still observe good
overall invisibility property except in the forward-scattering
direction.

The present cloak can be realized experimentally by designing proper
metamaterial structures. To achieve the constant $\varepsilon_{r}$,
we can use the electric resonant structures which align in the
\emph{r}-\emph{z} plane. To realize the varying $\mu_{z}$, we can
use the split-ring resonators which align in the \emph{r}-$\phi$
plane. The constant $\varepsilon_{\phi}=6.641$ can be obtained using
non-resonant structures like wires or I-shapes. It will be a hard
work to optimize the overall design to make a compact layout of
cylindrical cloak.

Compared with the full-parameter cloak [1], the advantage of the
proposed invisibility cloak is that none of the parameters is
singular and the changing range of all parameters is relatively
small. To compare with the reduced-parameter cloak [2], we compute
the scattering width for the PEC target only and the PEC target
covered by the reduced cloak [2], the proposed layered cloak without
loss, and the layered cloak with electric and magnetic-loss tangents
of 0.03, as shown in Fig. 3(c). Clearly, the proposed cloak has a
much better overall performance of invisibility.

This work was supported in part by the National Science Foundation
of China under Grant Nos. 60871016, 60671015, 60601002, and
60621002, in part by the Natural Science Foundation of Jiangsu
Province under Grant No. BK2008031, in part by the National Basic
Research Program (973) of China under Grant No. 2004CB719802, and in
part by the 111 Project under Grant No. 111-2-05.

\newpage

\section*{{{\bf List of Figure Captions}}}

\noindent \textbf{Fig. 1:} {\small (color online) An elliptical
cloak in the elliptical coordinate system. The blue (solid) and
green (dashed) lines represent constant $\xi$ and $\eta$ contours,
respectively. The cloaked object is crushed to a line segment $2p$.}

\vskip 5mm

\noindent \textbf{Fig. 2:} {\small (color online) (a) The
distributions of magnetic fields inside circular cloaks with
$R_1=0.025$ m, $R_2=0.05$ m, and $p=0.001$ m. (b) The distributions
of $\varepsilon_{\xi}$, $\varepsilon_{\eta}$, and $\mu_{z}$
components in the cloak region.}

\vskip 5mm

\noindent \textbf{Fig. 3:} {\small (color online) (a) The full-wave
simulation results for a layered lossless homogeneous cloak, in
which $R_1=0.025$ m, $R_2=0.05$ m. (b) The full-wave simulation
results for the layered homogeneous cloak with electric and
magnetic-loss tangents of 0.03. (c) Scattering width for the PEC
cylinder only, the PEC cylinder with the reduced-parameter cloak
[2], the layered homogeneous cloak, and the lossy layered cloak.}


\newpage

\begin{figure}
\centerline{\includegraphics[width=8cm,height=6cm]{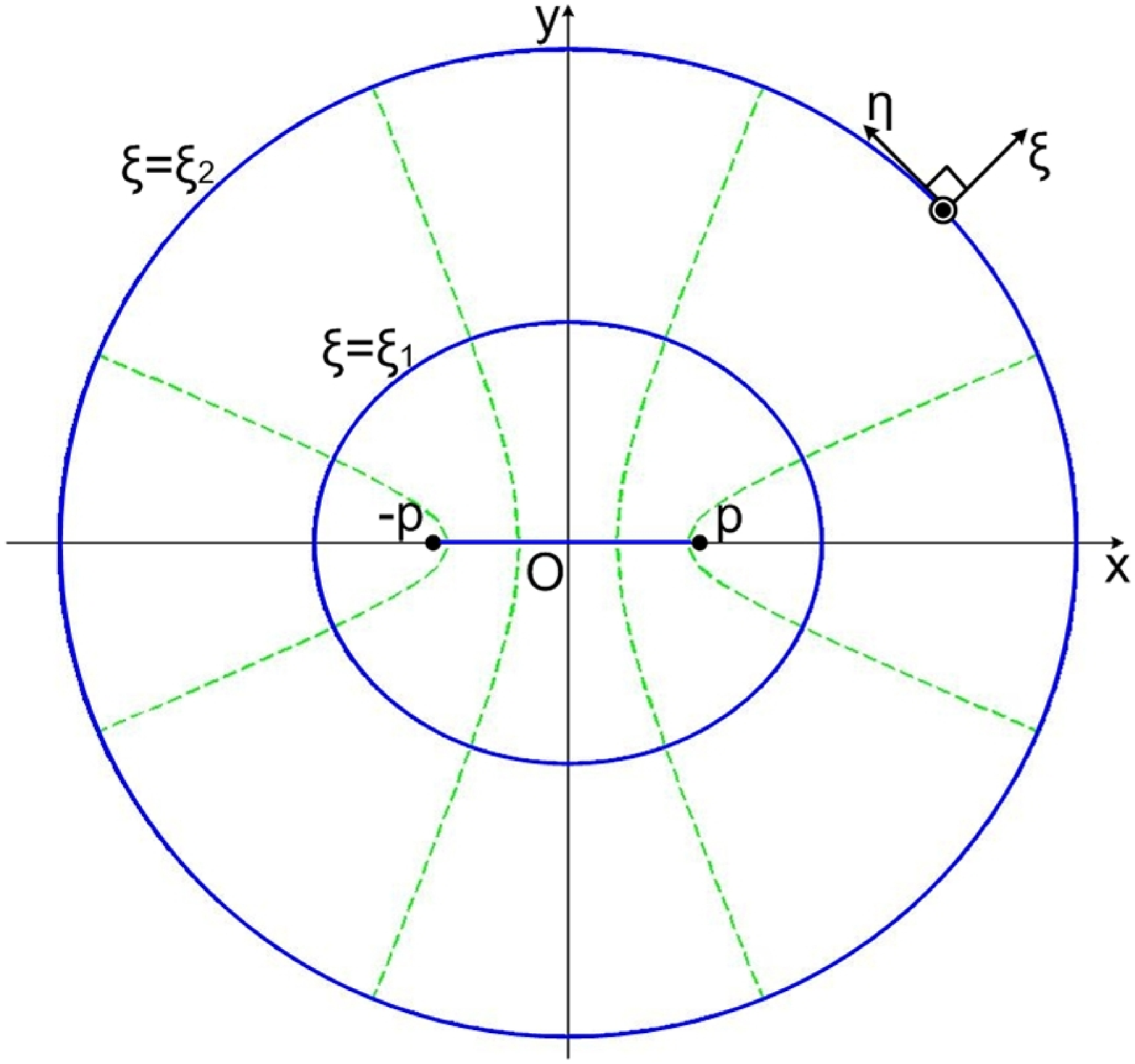}}
\caption{}
\end{figure}

\newpage

\begin{figure}
\centerline{\includegraphics[width=8cm]{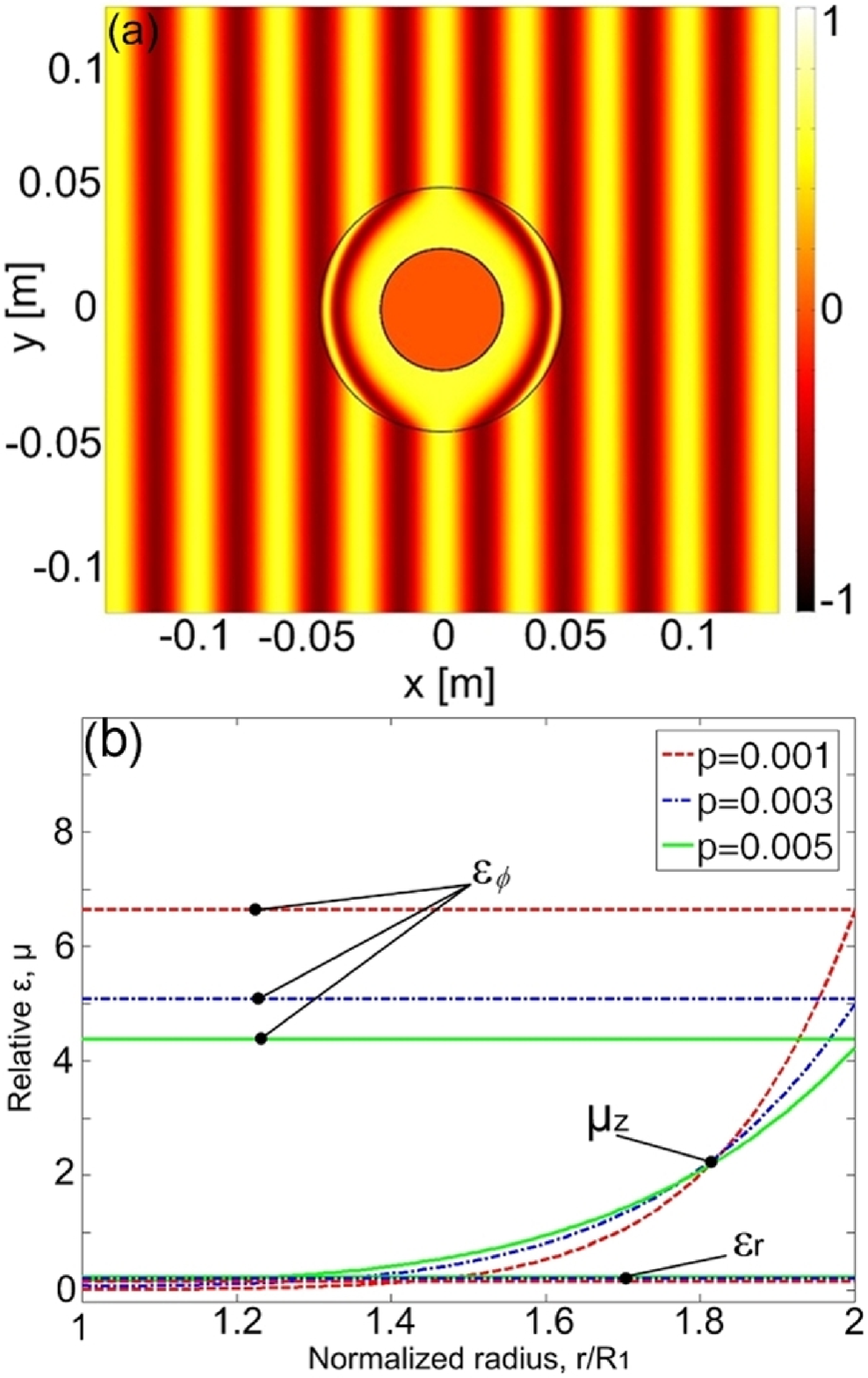}}
 \caption{}
\end{figure}

\newpage

\begin{figure}
\centerline{\includegraphics[width=8cm]{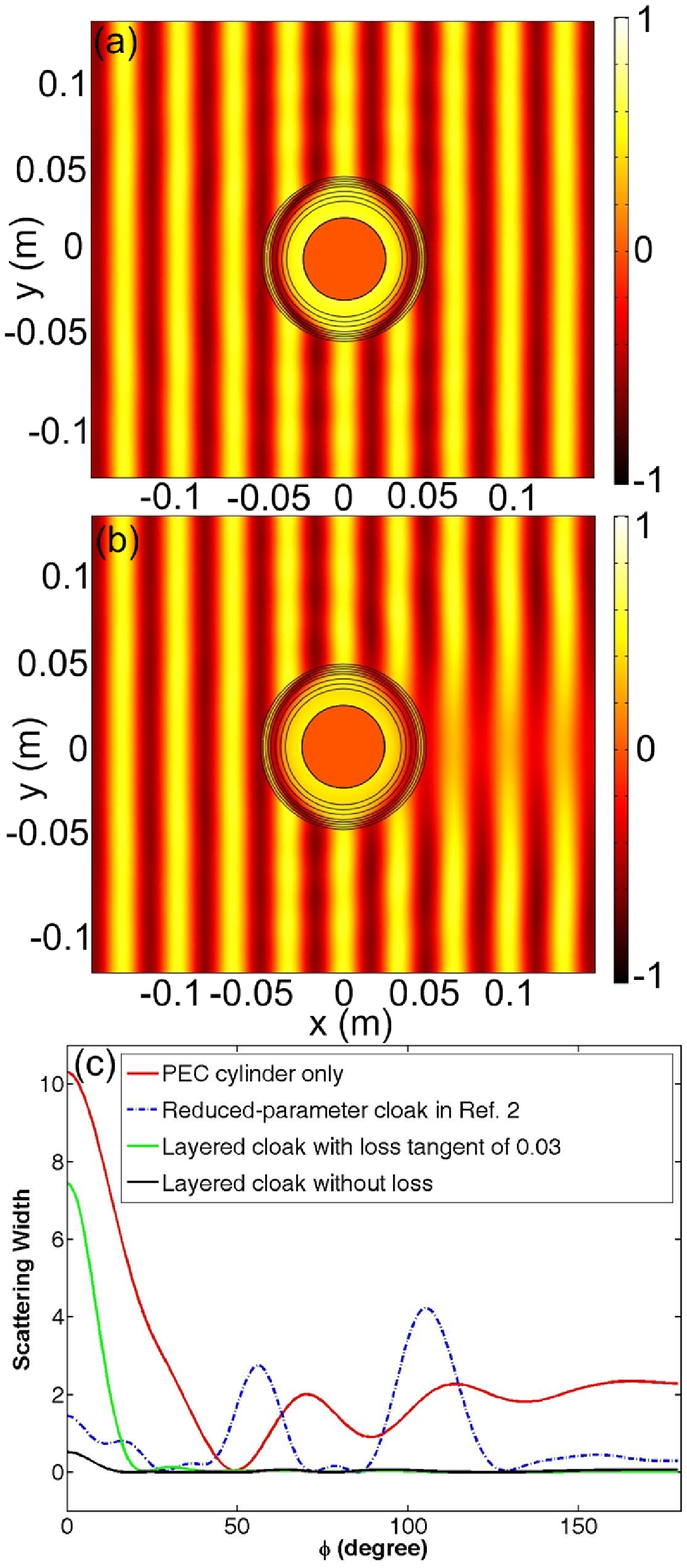}}
 \caption{}
\end{figure}

\end{document}